\documentclass[aps,pra,reprint,showpacs,superscriptaddress,longbibliography]{revtex4-1}
\usepackage[utf8]{inputenc}
\usepackage{xcolor}

\begin{document}
\title{Comment on the Quantum Supremacy Claim by Google}
\author{Anirudh Reddy }
\email{segireddya@ukzn.ac.za}
\affiliation{School of Chemistry and Physics, University of KwaZulu-Natal, Durban, South Africa}
\author{Benjamin Perez-Garcia}
\email{b.pegar@tec.mx}
\affiliation{Photonics and Mathematical Optics Group, Tecnologico de Monterrey, Monterrey 64849, Mexico}
\author{Adenilton Jose da Silva }
\email{ajsilva@cin.ufpe.br }
\affiliation{Centro de Inform\'atica, Universidade Federal de Pernambuco, Recife, Pernambuco, Brazil}
\author{Thomas  Konrad }
\email{konradt@uzkn.ac.za}
\affiliation{School of Chemistry and Physics, University of KwaZulu-Natal, Durban, South Africa}
\affiliation{National Institute of Theoretical and Computational Sciences (NITheCS), KwaZulu-Natal, South Africa}

\maketitle

Quantum computation promises to execute certain computational tasks on time scales much faster than any known algorithm on an existing classical computer, for example calculating the prime factors of large integers \cite{Shor1999}. Recently a research team from Google claimed to have carried out such a task with a quantum computer \cite{Google2019}, demonstrating in practice a case of this so-called quantum supremacy \cite{Preskill2018}.  Here we argue that this claim was not justified. Unlike other comments \cite{pednault2019leveraging, svozil2019comment, alicki2020meaning, Noh2020efficientclassical, barak2020spoofing, huang2020classical, Aaronson2020, pan2021simulating}, our criticism is concerned with the missing verification of the output data of the quantum computation.   \\

According to the description of the corresponding experiment, random quantum circuits were executed on a quantum computer to generate random numbers that cannot be produced in any feasible time
with known classical algorithms \cite{Google2019}. In the absence of noise in the quantum device, the occurrence of probabilities $P$ for random numbers with $n$ bits would obey a so-called Porter-Thomas distribution with expectation value $\langle P\rangle_{\rm \tiny best}= 2/2^n$ \cite{Google2019}.   Sampling such ideal probabilities would demonstrate quantum supremacy. However, on a scale with increasing white noise all outputs become eventually equiprobable with $\langle P\rangle_{\rm \tiny worst}= 1/2^n$ and thus reproducible with a classical computer or even manually by simply tossing $n$ coins with equal chance for heads and tails. 
    
For the purpose of verifying the results, a fidelity measure $F$ can be defined, that compares the average ideal probability of all outputs generated by the quantum computer, $\bar{P}_{\rm \tiny actual} \approx \langle P\rangle_{\rm \tiny actual} $, 
to the corresponding expectation values for a noiseless and a maximally noisy quantum computer.   Such a measure is given by the linear cross-entropy benchmarking fidelity $F$ \cite{Google2019, Neill2018, Boixo2017}, which  quantifies the performance of the quantum computer with values between $0$ and $1$, where one (zero) marks the best (worst) sampling of the desired probability distribution, 
$$ F= \frac{ \langle P\rangle_{\rm \tiny actual} - \langle P\rangle_{\rm \tiny worst}}{ \langle P\rangle_{\rm \tiny best} - \langle P\rangle_{\rm \tiny worst}} = 2^n  \langle P\rangle_{\rm \tiny actual} - 1. $$
At the same time this fidelity yields the success probability that the quantum computer produced the correct output states for its quantum bits, which was estimated to be of the order of magnitude $0.1\%$.

It would be too time consuming to measure the probabilities of the $N=2^{53}\approx 9\times10^{15}$ random numbers that Google's Sycamore quantum processor can generate.  
For this reason, fidelity $F$ appears to be a suitable figure of merit, since it deals with ideal probabilities that were not detected but calculated for each experimentally obtained output, based on the random circuit used, assuming its noiseless execution. Subsequently the ideal probabilities of all random numbers generated with random quantum circuits were averaged to approximate the value of $\langle P\rangle_{\rm \tiny actual}$.  In order to calculate the probabilities, classical super computers were employed. However, this could only be done for quantum circuits with small depth, not in the supremacy regime where the generation of the random numbers, let alone the more complex calculation of their probabilities, would not be possible on classical computers with known classical algorithms in any feasible amount of time. Hence, this method is not applicable in the supremacy regime, and consequently a verification of the set of random numbers produced in this regime is missing. 

Instead, the value of fidelity of the quantum computer in the supremacy regime was estimated based on a noise model and 
results for simplified quantum circuits of full depth that could be classically simulated. For circuits with small depth the theoretical fidelity value and the value obtained for the simplified circuits coincided fairly well with the fidelity of the actual circuits. Therefore, we can assume, and even conclude, that these fidelity values coincide in the supremacy regime as well, {\it if} the quantum computer works (with the same accuracy) in this regime. However, all values theoretically predicted or measured with simplified circuits do not imply {\it that} the quantum computer works (with the same accuracy) in the supremacy regime. For example, it is not implausible that the increase in complexity for full depth could decrease the performance to such a level that it is not superior to the performance of classical computers.

Indeed, if the value of fidelity $F$ is too low, the output data of Google's quantum computer could be simulated by a classical computer \cite{Google2019}, as is evident for $F=0$.  Hence, in the absence of a detected fidelity value, Google's experiment does not count as experimental proof of quantum supremacy. On the other hand, where it is possible to detect the fidelity values based on the calculation of the probabilities on a classical device, the quantum computer can be simulated and thus is not working in the supremacy regime. This refutes the claim of an experimental proof of quantum supremacy in \cite{Google2019}.

%

\end{document}